\documentstyle[12pt]{article}

\newcommand{\ct}[1]{\cite{#1}}
\newcommand{\bi}[1]{\bibitem{#1}}
\newcommand{\lleq}[1]{\label{eq:#1}}
\newcommand{\beql}[1]{\begin{equation} \lleq{#1}}
\newcommand{\req}[1]{(\ref{eq:#1})}
\newcommand{\bm}[1]{\mbox{\boldmath $#1$}}
\def\beq{\begin{equation}}     \def\eeq{\end{equation}}
                  
\def\beqa{\begin{eqnarray}}    \def\eeqa{\end{eqnarray}}
\def\ov{\over}
\def\lg{\left\langle}   \def\rg{\right\rangle}
   
\def\non{\nonumber}
\def\pt{p_\perp}
\def\Nev{N_{\rm ev}}
\def\sfd{{\sf d}}
\def\rmf{{\rm f}}

\begin{document}


\newcommand{\regpreprint}[4]
{\noindent\begin{minipage}[t]{\textwidth}\begin{center}
\framebox[\textwidth]
{$\rule[6mm]{0mm}{0mm}$ \raisebox{1.3mm}
{#1{Institut f\"ur Theoretische Physik der Universit\"at Regensburg}}
}

\vspace{2mm}    \rule{\textwidth}{0.2mm}\\
\vspace{-4mm}   \rule{\textwidth}{1pt}
\mbox{ }    #2    \hfill    #3   \mbox{ }\\
\vspace{-2mm}   \rule{\textwidth}{1pt}\\
\vspace{-4.2mm} \rule{\textwidth}{0.2mm}
\end{center}
\end{minipage}

\vspace{2mm}
\mbox{}\hfill AZPH--TH/92--45 \mbox{ }\\
\mbox{}\hfill HEPHY--PUB--575/92 \mbox{ }\\

\vspace{1mm}
\begin{center} {#4}  \end{center}
}   

\regpreprint{\large\bf}{March 1993}{TPR--92--47}
{\large\bf
Integral correlation measures for multiparticle physics}

\vspace{5mm}
\begin{center}
H.\ C.\ Eggers\footnote{eggers@rphs1.physik.uni-regensburg.de}\\
Institut f\"ur Theoretische Physik, Universit\"at Regensburg,\\
D--93040 Regensburg, Germany \\
\mbox{ }\\
\mbox{ }\\
P.\ Lipa\footnote{lipa@arizvm1.ccit.arizona.edu},
P.\ Carruthers\footnote{carruthers@ccit.arizona.edu}\\
Department of Physics, University of Arizona, Tucson AZ 85721, USA\\
\mbox{ }\\
\mbox{ }\\
B.\ Buschbeck\footnote{v2033dae@awiuni11.edvz.univie.ac.at}\\
Institut f\"ur Hochenergiephysik\\
\"Osterreichische Akademie der Wissenschaften\\
Nikolsdorfergasse 18, A--1050 Vienna, Austria
\end{center}

\mbox{}\\

\begin{abstract}
We report on a considerable improvement in the technique of
measuring multiparticle correlations via integrals over correlation
functions. A modification of measures used in the characterization
of chaotic dynamical sytems permits fast and flexible calculation of
factorial moments and cumulants as well as their differential
versions. Higher order correlation integral measurements even of large
multiplicity events such as encountered in heavy ion collisons are now
feasible. The change from ``ordinary'' to ``factorial'' powers may have
important consequences in other fields such as the study of galaxy
correlations and Bose-Einstein interferometry.
\end{abstract}

\begin{center}
\mbox{ }\\
submitted to Physical Review D
\end{center}

\newpage

\section{Introduction}

Wherever statistical analyses are done, whether in physics,
biology or psychology, the measurement of the correlation
function is a basic element of understanding. While each
discipline has its own set of questions for which it seeks
answers, the underlying statistical mechanisms are very
similar: given a set of variables, one first finds the
distribution of how often each of these variables takes
on a certain value, and  much that is useful can be learned
from these one-variable distributions. Following immediately
is the question how two variables behave simultaneously, whether
they are independent or in some way correlated: the two-variable
correlation function provides the answers. Higher orders provide
additional information, but with escalating difficulty of
measurement and diminishing returns.

Conversely, a knowledge of correlation functions to all orders
provides complete information on any statistical system.

Of special interest to us here are point distributions.
Typical examples of point distributions include galaxies in the
sky, cows in a field and particles in phase space (the exact size
of the object under study is irrelevant as long as it is small
compared to the embedding space). The aim of this paper
is to develop and extend methods of measuring correlation
functions of point distributions. While we shall be
considering particle correlations in
high energy collisions, the formalism developed here should
be suitable, with appropriate modifications,
for problems in a number of other situations.

Beyond traditional methods,
recent advances in the theory of fractals and
scaling have spawned a new way of approaching correlations:
scaling behavior manifests itself in power-law behavior
of the correlation function, which in particle physics can be
measured experimentally through
the factorial moments revived by Bia\l as and Peschanski
\ct{Bia86a}. On the other hand, a measure termed the correlation
integral \ct{GHP83,Pal84a,Pal87a} has been in use in the
characterization of strange
attractors and other contexts for some time.
As an improvement on the factorial moments and following the
suggestion of Dremin \ct{Dre88a}, we previously
introduced into multiparticle physics two forms of the correlation
integral which we termed the ``Snake'' and the
symmetrized ``GHP'' integrals \ct{Lip92a}.
In this paper, we advocate a slightly different form,
which for obvious reasons we name ``Star'' integral.
Yet another form useful for pion interferometry takes the
invariant mass of the $q$-tuple as a
measure of its size \ct{Jur89a,Egg93d}.

To illustrate the concept of the correlation integral and the
difference between the various forms, we consider
the phase space plot (e.g.\ in rapidity and azimuthal angle)
of the pions produced in a particular collision and ask the
question how clusters consisting of $q$ particles
may be characterized, i.e.\ what ``size'' should be assigned to each
cluster. Taking $q=5$, for example, we show in
Figure 1 one particular selection of five particles from
this event and assign the following sizes to it: in (a), the five
particles are joined into a ``snake'', and the 5-tuple is assigned a
size $\epsilon$ corresponding to the longest
of the four joining lines. In (b), the same 5-tuple
is assigned a size given by the maximum of all ten
pairwise distances; this defines the ``GHP'' correlation integral.
The ``Star'' integral in Figure 1(c), finally, is assigned a
size corresponding to the longest of the four lines to the
chosen center particle.
A line represents a particular interparticle distance being
{\it tested} and found to be smaller than $\epsilon$;
those particles {\it not} connected by a line may or may not
be within a distance $\epsilon$ of each other.

Every correlation integral of order $q$ thus assigns a size
to every possible $q$-tuple of particles in an event. The way
this assignment is done distinguishes the different versions of
correlation integrals. Once such an assignment
is made, they all count the
number of $q$-tuples that are smaller than a given size.
For a large data sample this corresponds to an integration over the
$q$-th order correlation function, as shown
explicitly in Eqs.\ \req{ciga}--\req{cigc} below.

We shall show that
the Star shares all the advantages of the Snake and
GHP forms but is more amenable to intuitive understanding and is
computationally more efficient by orders of magnitude.
Furthermore, our formalism leads naturally to a conceptual cleanup
of the heuristic fractal measure used in the study
of strange attractors, in galaxy distributions and in multiparticle
correlations  \ct{Pal84a,Pal87a,Dre88a,Lip92a,Col92a}.
Being derived
directly and rigorously from the underlying correlation function,
it necessitates the use of {\it factorial\/}
powers rather than the ordinary powers used previously.
The difference becomes significant when small
particle numbers are involved, a situation which occurs
inevitably when the integration domain becomes sufficiently small.

Besides this cleanup, we provide here a technique to
measure integrals of cumulant correlation functions which
are the genuine higher order correlations and thus very useful
observables. Previously, the only way to
extract information on higher order cumulants had been via
combinations of factorial moments \ct{Car90d} and in third order
for some very special configurations in rapidity space \ct{ABCDHW}.
By clarifying and
extending the technique of event mixing, we show how
cumulant functions can be integrated directly and thus share
all the advantages of correlation integrals over
bin-averaged factorial moments.

The paper is organized as follows. In Section \ref{sec:cori},
the Star correlation integral is introduced, while in
Section \ref{sec:evmix} we lay a solid
theoretical foundation for the practice of event mixing.
In Section \ref{sec:cum},
we remind the reader of the basic relations defining true
correlations and show how these may be measured directly in the
correlation integral scheme. Differential versions and their
considerable advantages are discussed in Section \ref{sec:diff}.
We conclude with some remarks and recommendations.

\section{The correlation integral}
\label{sec:cori}

\subsection{Basic concepts}

The starting point for all correlation analyses is the
$q$-th order correlation function or ``inclusive density''
$\rho_q(\bm{x}_1,\bm{x}_2,\ldots,\bm{x}_q)$.
The choice of the $q$ variables $\bm{x}_k$ living in $\sf d$
dimensions is determined by our particular problem;
in high energy physics, we can have, for example,
vector three-momenta
$\bm{x} \equiv (p_x,p_y,p_z)$, some combination
of boost-invariant variables such as rapidity, azimuthal angle
and transverse momentum $(y,\phi,\pt)$
or just one of these alone\footnote{
In addition to a suitable choice of variables, some
``pre-processing'' of the data may be required to eliminate
unwanted effects, for example the transformation to jet
coordinates in $e^+e^-$ collisions \ct{Aih85a}, and, if desired,
the creation of data subsamples of fixed total multiplicity.
}.

The set of correlation functions $\rho_q,\ q = 1,2,\ldots$ can
be defined for a fixed total number of particles and/or for
specific particle types such as positively and negatively
charged pions; for the purposes of this paper, we shall mostly
consider
only one type of particle within an inclusive distribution where
the total number of particles per event is not held constant.
For this case, $\rho_q$ is defined operationally as
\beql{cic}
\rho_q(\bm{x}_1,\ldots,\bm{x}_q) = {1 \ov \sigma_I}
{ d^q\sigma_{\rm incl} \ov
d\bm{x}_1\,d\bm{x}_2\ldots d\bm{x}_q} \;,
\eeq
where $\sigma_I$ is the total interaction cross section and
$\sigma_{\rm incl}$ the inclusive cross section.
Integrating in $\sf d$ dimensions over the total window
$\Delta \bm{x} = (\Delta x)^{\sf d}$, we get
\beql{cie}
\int_{\Delta \bm{x}}
d\bm{x}_1\ldots d\bm{x}_q\, \rho_q(\bm{x}_1,\ldots,\bm{x}_q) =
\lg N^{[q]} \rg \ ,
\eeq
where $N$ is the total number of particles in
$\Delta \bm{x}$ and
$\lg N^{[q]}\rg = \lg N(N{-}1)\ldots(N{-}q{+}1) \rg
= \xi_q(\Delta x)$
is the unnormalized $q$-th order factorial moment over
the same region.

Integrating $\rho_q$ over various domains of integration,
one can obtain any number of possible moments. For example, the
vertical normalized factorial moment revived by Bia\l as and
Peschanski \ct{Bia86a},
\beql{cif}
F_q^v \equiv {1\ov M^{\sfd}}
\sum_{m_1,\ldots,m_{d}=1}^M
{
\lg n_{m_1,\ldots,m_{d}}^{[q]} \rg
\ov
\lg n_{m_1,\ldots,m_{d}} \rg^{q}
}
=
{1\ov M^{\sfd}}
\sum_{m_1,\ldots,m_{d}=1}^M
{
\int_{\Omega({\sf m)}} \prod_k  d\bm{x}_k \,
\rho_q(\bm{x}_1,\ldots ,\bm{x}_q)
\ov
\int_{\Omega({\sf m)}} \prod_k  d\bm{x}_k \,
\rho_1(\bm{x}_1)\cdots\rho_1(\bm{x}_q)
}\,,
\eeq
integrates $\rho_q$ in a cartesian lattice
of $M^{\sfd}$ cubes, each of size
$\Omega({\sf m}) = \Omega(m_1,\ldots,m_{\sfd}) = (\delta x)^{\sfd}$,
normalizing each bin separately. For the purposes of searching
for power-law behavior of the correlation function, $F_q$
is measured as a function of $M$, with the bin edge length
decreasing correspondingly, $\delta x = \Delta x/M$. The
``horizontal'' factorial moments \ct{Bia86a}, the differential
versions of Section \ref{sec:diff} and indeed the
traditional way of presenting correlation functions in constant
bin sizes are all integrations over different domains of the
same correlation function.

In the same way, the three correlation integrals are simply
integrals over specific domains.
The $q$-th order Snake correlation integral is defined
in terms of $\rho_q$ and the ``cluster size'' $\epsilon$ as
\beql{ciga}
F_q^{\rm Snake}(\epsilon) \equiv
{
\int \rho_q(\bm{x}_1,\ldots,\bm{x}_q) \,
\Theta_{12}\Theta_{23}\ldots\Theta_{q-1,q} \,
d\bm{x}_1 \ldots d\bm{x}_q
\ov
\int \rho_1(\bm{x}_1)\ldots\rho_1(\bm{x}_q) \,
\Theta_{12}\Theta_{23}\ldots\Theta_{q-1,q} \,
d\bm{x}_1 \ldots d\bm{x}_q
} ,
\eeq
where $\Theta_{ij} \equiv \Theta(\epsilon - |\bm{x}_i - \bm{x}_j|)$.
Similarly, the GHP moment is defined with all interparticle
distances restricted\footnote{
In Ref.\ \ct{Lip92a}, we erroneously denied that the GHP
integral could be written down as an analytical integral of
the correlation function.},
\beql{cigb}
F_q^{\rm GHP}(\epsilon) \equiv
{
\int \rho_q(\bm{x}_1,\ldots,\bm{x}_q) \,
\prod_{i<j} \Theta_{ij}\,
d\bm{x}_1 \ldots d\bm{x}_q
\ov
\int \rho_1(\bm{x}_1)\ldots\rho_1(\bm{x}_q) \,
\prod_{i<j} \Theta_{ij}\,
d\bm{x}_1 \ldots d\bm{x}_q
} ,
\eeq
where
\beql{cigbb}
\prod_{i<j = 1}^q \Theta_{ij} =
\Theta_{12}\Theta_{13}\ldots\Theta_{1q}\Theta_{23}\ldots\Theta_{q-1,q}
\eeq
restricts all possible pairs of coordinates.
The Star integral is defined as
\beql{cigc}
F_q^{\rm Star}(\epsilon) \equiv
{
\int \rho_q(\bm{x}_1,\ldots,\bm{x}_q) \,
\Theta_{12}\Theta_{13}\ldots\Theta_{1q} \,
d\bm{x}_1 \ldots d\bm{x}_q
\ov
\int \rho_1(\bm{x}_1)\ldots\rho_1(\bm{x}_q) \,
\Theta_{12}\Theta_{13}\ldots\Theta_{1q} \,
d\bm{x}_1 \ldots d\bm{x}_q
} ,
\eeq
involving all coordinates paired with $\bm{x}_1$.
The ``topology'' of the different correlation integrals shown in
Figure 1 is thus visible already in the selection of
theta functions.

\subsection{The Star integral}
\label{sec:sta}

While the definition \req{cic} of $\rho_q$ is exact and
unambiguous, we shall need for the derivation of the
correlation integral an alternative but equivalent formulation
written down by e.g.\ Klimontovich \ct{Kli67a}.
Let the $N$ particles of a specific event be situated at the points
$\bm{X}_i$ in phase space, $i = 1,\ldots,N$. Then the ``event
correlation function'' $\hat\rho_q$ is defined as adding 1
at every point $(\bm{x}_1,\ldots,\bm{x}_q)$ if there is
simultaneously at each $\bm{x}_k$ a particle $\bm{X}$,
independently of the positions of the other $N-q$ particles.
This is done for all $N^{[q]} = N!/(N-q)!$
selections of $q$-tuples out of the total $N$ particles.
For a specific event $a$ this defines a function
\beql{ciaa}
\hat\rho_q^a(\bm{x}_1,\ldots,\bm{x}_q) =
\sum_{i_1\neq i_2\neq\ldots\neq i_q \atop =1}^{N}
\delta (\bm{x}_1-\bm{X}^a_{i_1})\,
\delta (\bm{x}_2-\bm{X}^a_{i_2})    \,\cdots\,
\delta (\bm{x}_q-\bm{X}^a_{i_q}) \,,
\eeq
where $\delta(\bm{x})$ is the product of $\sf d$ one-dimensional
delta functions.
This function, when averaged over all events, yields the
$q$-particle distribution function of Eq.\ \req{cic},
\beql{cid}
\rho_q(\bm{x}_1,\ldots,\bm{x}_q)
=  \lg  \hat\rho_q^a(\bm{x}_1,\ldots,\bm{x}_q) \rg
= \Nev^{-1} \sum_{a=1}^{\Nev}  \hat\rho_q^a(\bm{x}_1,\ldots,\bm{x}_q)
\,,
\eeq
where $\Nev$ is the number of events in the experimental sample.
For finite resolution, Eq.\req{cid} corresponds to building up a
$q{\sfd}$-dimensional histogram\footnote{
Since both definitions \req{cic} and \req{cid} are implemented
for a finite number of events, they are, strictly speaking,
estimators of the true correlations.
}.

Inserting Eqs.\ \req{ciaa}--\req{cid} into the numerator of
Eq.\ \req{cigc},  we find immediately the (unnormalized)
Star integral factorial moment to be
\beql{ciab}
\xi_q^{\rm Star}(\epsilon) = \left\langle
\sum_{i_1\neq i_2\neq\ldots\neq i_q}^{N}
\Theta(\epsilon- X_{i_1 i_2})
\Theta(\epsilon- X_{i_1 i_3}) \ldots
\Theta(\epsilon- X_{i_1 i_q})
\right\rangle ,
\eeq
where $X_{i_1 i_k} = |\bm{X}_{i_1} {-} \bm{X}_{i_k}|$. Pulling out
the first sum, we can factorize the remaining sums,
\beql{ciac}
\xi_q^{\rm Star}(\epsilon) = \left\langle
\sum_{i_1}
\left( \sum_{i_2 \neq i_1}
\Theta(\epsilon- X_{i_1 i_2})
\right)^{[q-1]} \right\rangle  ,
\eeq
where the factorial power $[q{-}1]$ in the exponent came about
because the sum indices are restricted to
$i_\alpha \neq i_\beta \neq i_1$
for all $\alpha \neq \beta$. The quantity inside
the round brackets is so important that we give it a special name,
the {\it sphere count}:
\beql{ciad}
\hat n(\bm{X}_{i_1},\epsilon)
\equiv
\sum_{i_2=1}^N \Theta(\epsilon - X_{i_1 i_2})
\,,\ \ \ \ \ \ i_2 \neq i_1 \,,
\eeq
which represents the number of
particles within a sphere of radius $\epsilon$ centered on the
particle $\bm{X}_{i_1}$, excluding the center particle itself.
(For the given $\epsilon$ and center particle shown in Figure
2(a), we would have $\hat n(\bm{X}_{i_1},\epsilon) = 9$.)
In a similar
form, it is used extensively in the description of galaxy
distributions. Introducing the shorthand notation
${\hat n}(\bm{X}_{i_1}^a,\epsilon) \equiv a$
the unnormalized factorial moment can be written compactly as
(we henceforth drop the ``Star'' superscript)
\beql{ciae}
\xi_q(\epsilon) =
\Nev^{-1} \sum_a \sum_{i_1}
         \hat n(\bm{X}_{i_1}^a,\epsilon)^{[q-1]}
\equiv
\lg \sum_{i_1} a^{[q-1]} \rg \,.
\eeq

An alternative derivation of $\xi_q(\epsilon)$ proceeds by
coordinate transformation \ct{Lip92a,Car91ap}.
One first defines a ``particle-centered'' correlation function
around the particle at $\bm{X}_{i_1}$, fixing to it the
coordinate $\bm{x}_1$ by a delta function,
\beql{ciaf}
\hat\rho_q(\bm{X}_{i_1}
     ;\bm{x}_1,\bm{x}_2,\ldots,\bm{x}_q)
= \delta(\bm{x}_1 - \bm{X}_{i_1})
\sum_{i_2\neq i_3 \neq \ldots \neq i_q}^N
\delta(\bm{x}_2 - \bm{X}_{i_2})
\delta(\bm{x}_3 - \bm{X}_{i_3})  \,\cdots\,
\delta(\bm{x}_q - \bm{X}_{i_q})  \,,
\eeq
where the sum indices are all restricted additionally by
$i_\alpha \neq i_1$, and then transforms to relative coordinates.
These are the distinctive hallmark of correlation integrals:
for the Snake integral, we used the coordinate transformation
$R = \sum_{k=1}^q x_k/q$ and $r_k = x_{k+1} - x_k$ \ct{Lip92a}.
For the Star integral, on the other hand, all coordinates are
defined relative to $\bm{x}_1$,
\beqa
\lleq{cih}
\bm{r}_1 = \bm{R} &=& \bm{x}_1 \,, \non\\
         \bm{r}_k &=& \bm{x}_{k} - \bm{x}_1 \,,
\ \ \ \ \ \ k = 2,\ldots,q \,.
\eeqa
Inserting these into the delta functions of Eq.\req{ciaf}, we find
\beqa
\lleq{cii}
\hat\rho_q(\bm{X}_{i_1};\bm{R},\bm{r}_2,\ldots,\bm{r}_{q})
&=&  \delta[\bm{R}       -  \bm{X}_{i_1}   ]
   \sum_{i_2\ne i_3\ne \ldots\ne i_q}
   \delta[\bm{r}_2  - ( \bm{X}_{i_2} - \bm{X}_{i_1}    )]
        \times \ldots
\nonumber \\
& &\ \ \ldots\times
   \delta[\bm{r}_q  - ( \bm{X}_{i_q} - \bm{X}_{i_1}  )  ]  \;.
\eeqa
Since at this point we are only interested in correlations as
a function of relative distances, we integrate out
$\bm{R}$ over the entire interval
$\Delta\bm{R} = \Delta\bm{x}$.
The relative coordinates $\bm{r}_k$ we want to restrict to
a maximum length $r_k = |\bm{r}_k| \leq \epsilon$.
For higher dimensions ${\sf d} > 1$, we must first integrate out
the angular parts $d\Omega_k$ of $d\bm{r}_k$.
Since we shall eventually normalize our correlations using
exactly the same domain of integration, however,
the constants resulting from the angular integrations will
cancel and we henceforth ignore them. The remaining
integral over the lengths $r_k$ is given by
$\int_0^\epsilon \prod_k\, dr_k\, r_k^{{\sf d}-1}$.
At the same time, on integrating out the angular coordinates,
the remaining delta functions of Eq.\ \req{cii} become
\beql{cij}
\int d\Omega_k \, \delta[\bm{r}_k - \bm{X}_{i_1 i_k}]
=
{\delta[r_k - X_{i_1 i_k}] \ov X_{i_1 i_k}^{{\sf d}-1} }\,,
\eeq
and the factors $r_k^{{\sf d}-1}$ will on integration cancel exactly
with $X_{i_1 i_k}^{{\sf d}-1}$. To express this entire process of
simplification, we shall write in shorthand
\beql{cik}
\int_0^\epsilon \prod_k d\bm{r}_k \,,
\eeq
which is just an integral of the lengths $r_k$.
For ${\sf d} = 1$,
$\int_0^\epsilon \prod_k d\bm{r}_k$ is shorthand for
$\int_{-\epsilon}^\epsilon \prod_k dr_k$.

Integrating in this way over all relative coordinates, we again get
a factorial product of single sums:
\beqa
\lleq{cila}
& &
\int_{\Delta\bm{x}} d\bm{R}
\int_0^\epsilon \prod_{k=2}^q d\bm{r}_k \,
\hat\rho_q(\bm{X}_{i_1};\bm{R},\bm{r}_2,
                        \ldots,\bm{r}_q) \non\\
&=&
\sum_{i_2 \neq \ldots\neq i_q}
\Theta(\epsilon - X_{i_1 i_2})
\Theta(\epsilon - X_{i_1 i_3}) \,\ldots\,
\Theta(\epsilon - X_{i_1 i_q})  \non\\
&=&
\left( \sum_{i_2} \Theta(\epsilon - X_{i_1 i_2} ) \right)^{[q-1]} ,
\eeqa
which, on summing over $i_1$ and averaging over all events,
again yields Eq.\ \req{ciac}.

As derived above, the
$\xi_q(\epsilon)$ of Eqs.\ \req{cigc} and
\req{ciac}--\req{ciae} is
the unnormalized factorial moment over the domain shown in
Fig.\ 2(a),
\beql{cilf}
\Omega_S(\epsilon) \equiv
\left\{ \bm{R},\bm{r}_k |
\bm{R} \in \Delta\bm{x}, r_k \in [0,\epsilon],
k = 2,\ldots,q \right\} \,;
\eeq
this ``Floating Sphere'' form is best whenever the nature of our
variables $\bm{x}$ in $\sf d$ dimensions is such that a length
can be sensibly defined, the most obvious being the euclidean
distance
\beql{ciq}
r_k \equiv \sqrt{r_{k,x}^2 + r_{k,y}^2 + \ldots} \,,
\eeq
but often this is not so good. The
variables $(y,\phi,p_\perp)$, for example, have very different behavior,
and it may be better to treat each one separately. For such cases,
one may use the ``Floating Box'' form \ct{Janpc},
where each of the $\sf d$
components is treated as a one-dimensional correlation integral
and the total domain as shown in Figure 2(b) is
\beql{ciqa}
\Omega_{B}(\epsilon) \equiv
\left\{ R_f, r_{k,f}
| R_f \in \Delta x, r_{k,f} \in [-\epsilon,\epsilon] \;
\forall  k = 2,\ldots,q; f = y,\phi,\ldots \right\} \,.
\eeq
This corresponds to inserting ${\sf d}(q{-}1)$ theta functions
into Eq.\ \req{cigc}, one for each component.

The results of this section have the following important consequences:
\begin{enumerate}
\item
As shown in Eqs.\ \req{cila}--\req{ciac}, the $q{-}1$ sums
factorize nicely into a single sum of sphere counts
$\hat n(\bm{X}_{i_1},\epsilon)$. This means that
$\xi_q(\epsilon)$ can be calculated in an algorithm of order
$N^2$, independently of the order $q$. As emphasized previously \ct{Pal84a},
this represents a tremendous savings in CPU time over other
correlation measures, including the Snake and GHP integrals
advocated by us previously \ct{Lip92a}, which run under
$N^q$ and $N^q/q!$ algorithms respectively.
\item
This means that the correlation integral can now be used also for
correlation analysis in heavy ion collisions, something hitherto
impossible due to the large event multiplicities.
The big improvements in statistics over the conventional vertical
factorial moments will allow for much more accurate measurements.
\item
Unlike the Star integral found so far in the literature
\ct{Pal84a,Pal87a,Dre88a,Atm89a},
we get a {\it factorial} power of the sphere count
${\hat n}^{[q-1]}$ rather than the ordinary power ${\hat n}^{q-1}$.
This result we obtained rigorously from first principles
merely by restricting the sum indices to be unequal because the
same particle may not be counted more than once. Even for the large
multiplicities encountered in astronomy, the change is not
inconsequential, as the difference between
${\hat n}^{[q-1]}$ and ${\hat n}^{q-1}$ is important when
$\epsilon$ becomes sufficiently small.
\item
We obtained this factorial power without drawing on distinctions
between ``dynamical'' and ``statistical'' fluctuations \ct{Bia86a}.
\item
To first order, we have ignored the variation of
$\xi_q(\epsilon)$ with the center coordinate
$\bm{R}$; this is equivalent to assuming that the
physics is the same for all parts of the defined window.
When statistics permit, it may be very useful to measure $\xi_q$ also
as a function of $\bm{R}$. For example, one expects the correlation
structure at small transverse momentum to be very different from
that at large $p_\perp$, so that a separate measurement of $\xi_q$ for
small and large $R \equiv p_{\perp 1}$ may be most revealing.
\item
The advantages of correlation integrals over the traditional
factorial moments arise because the former use
interparticle distances directly while the latter rely on fixed
bins and grids rather than the particle positions themselves.
\item
We must emphasize that the Star integral as derived here is
different from the Snake and GHP versions we used in earlier papers.
All three are correlation integrals, but they differ in the topology
of the interparticle distances measured. For $q=2$, all three
are the same, while the Snake and Star versions are the same even
at the level $q=3$. Only in fourth order do the differences between
the latter two appear.
While the Snake and GHP
versions are not wrong and an improvement over previous work,
the present Star integral represents a big step forward.
\end{enumerate}

\newpage

\section{Normalization by event mixing}
\label{sec:evmix}

We now proceed to consider the denominator of Eq.\ \req{cigc}.
Since the topic of normalization is complex and full of pitfalls,
we do not address the full range of issues here and defer such
discussion to future work. Instead, we concentrate on the
normalization scheme we consider most suitable for the
measurement of correlations in high energy physics,
a version resembling the so-called
vertical normalization used in Eq.\ \req{cif}.

The normalization of the correlation integral is done by means
of {\it event mixing}, a seemingly heuristic technique commonly
used in Bose-Einstein correlations \ct{Aih85a}. It is well founded,
however, both for our purposes here and in the Bose-Einstein
context \ct{Egg93d}. We recall that if all $q$ coordinates
$\bm{x}_k$ are statistically independent of
each other, the correlation function
factorizes: $\rho_q = \rho_1^q$. We hence normalize the numerator
\req{ciac} of the Star integral by integrating
$\prod_{k=1}^q\; \rho_1(\bm{x}_k=\bm{R}{+}\bm{r}_k)$
over the same domain,
\beqa
\lleq{evb}
\xi_q^{\rm norm}(\epsilon) &\equiv&
\int_{\Delta \bm{x}} d\bm{R} \, \rho_1(\bm{R})
\int_0^\epsilon  \prod_{k=2}^q d\bm{r}_k \,
     \rho_1(\bm{R}{+}\bm{r}_2)\ldots
     \rho_1(\bm{R}{+}\bm{r}_q) \non\\
&=&
\int_{\Delta \bm{x}} d\bm{R} \, \rho_1(\bm{R})
\left[
\int_0^\epsilon  d\bm{r}_2 \,
     \rho_1(\bm{R}{+}\bm{r}_2)
\right]^{q-1} ,
\eeqa
(here and below it is understood that one transforms from
$\bm{x}$-  to $\bm{r}$-coordinates before
integration). Inserting
$\rho_1(\bm{x}_k) = \Nev^{-1}\sum_{e_k}\sum_{i_k}
\delta(\bm{x}_k - \bm{X}_{i_k}^{e_k})$
for each factor, we find after integration
\beql{evc}
\xi_q^{\rm norm}(\epsilon) =
\Nev^{-1} \sum_{e_1}\sum_{i_1} \left[
\Nev^{1-q} \sum_{e_2,\ldots,e_q} \sum_{i_2,\ldots,i_q}
\Theta(\epsilon - X_{i_1 i_2}^{e_1 e_2}) \cdots
\Theta(\epsilon - X_{i_1 i_q}^{e_1 e_q})
\right] ,
\eeq
where now $X_{i_1 i_k}^{e_1 e_k} \equiv
| \bm{X}_{i_1}^{e_1} - \bm{X}_{i_k}^{e_k} |$
measures the distance between two particles {\it taken from
different events} $e_1$ and $e_k$.
This much resembles the numerator expressions of Eq.\ \req{ciac},
and indeed the sums also factorize here, so that
\beql{evd}
\xi_q^{\rm norm}(\epsilon) =
\left\langle \sum_{i_1}
\left\langle \sum_{i_2}
\Theta(\epsilon - X_{i_1 i_2}^{e_1 e_2})
\right\rangle^{q-1}
\right\rangle
\equiv
\left\langle
\sum_i \langle \hat n_b(\bm{X}_i^a,\epsilon)\rangle^{q-1}
\right\rangle ,
\eeq
or, in shorthand,
\beql{eve}
\xi_q^{\rm norm}(\epsilon) =
\left\langle \sum_i \langle b \rangle^{q-1} \right\rangle .
\eeq
Comparing the numerator \req{ciac} to the denominator \req{eve},
we note that the exponent of the latter is an ordinary power
$q{-}1$ instead of the factorial power $[q{-}1]$ of the
former; this follows from the fact that in Eq.\ \req{evc}
there are no restrictions on the sum indices.

Figure 3 shows how the denominator sphere count
$\hat n_b(\bm{X}_i^a,\epsilon) \equiv b$
is found by inserting the particle $\bm{X}_i^a$ of
event $a$ into another event $b$ of the sample and doing the
count $\hat n_b(\bm{X}_i^a,\epsilon)
= \sum_j \Theta(\epsilon - X_{ij}^{ab})$ around it.
This is done for many events $b$ to obtain the inner event
average of Eq.\ \req{evd}.
Note that one has to distinguish carefully between $a$- and
$b$-event averages: the computation of
$\langle b \rangle =
\langle\hat n_b(\bm{X}_i^a,\epsilon) \rangle$
involves an average over different events $b$ {\it while the
position of the sphere center is kept fixed} at $\bm{X}^a_i$.
Expressions of the form
$\langle \sum_i \langle b \rangle\rangle$ then denote sums of
contributions when the center of the sphere ``floats'' over the
$i=1,\ldots,N$ particle positions of event $a$ and finally
over all events $a=1,\ldots,\Nev$.

While the derivation of Eq.\ \req{evd} from \req{evb} is exact,
this expression for the normalization is
correct only for $\Nev\to\infty$: it contains a hidden bias due to
correlations induced by use of the same events in every factor
$\rho_1$ in \req{evb}. For the case where the
the experimental sample size is not infinite, statistical
theory provides {\it estimators}, which from the limited-size sample
estimate quantities for the ``true'' (i.e. infinitely
large) sample. Applying the theory of estimators to our
problem, we find that the product of distributions $\rho_1$ in
Eq.\ \req{evb} must be modified precisely in such a way
that the event indices
$e_1,\ldots,e_q$ are all mutually unequal. This is in agreement
with the heuristic procedure of creating ``fake events'' where
each track is taken from a different (real) event.

The corrections to obtain such an {\it unbiased} form of the
normalization
can be written as a series in powers of $1/\Nev$, with the leading
term given by $\langle b \rangle^{q-1}$. For the relatively
small number of events and great sensitivity to bias found in heavy ion
samples, the correction can be quite important, while the
situation for hadronic data is less acute. We defer the details
of this very technical discussion and the exact expressions
to a future publication \ct{Liptbp}.

Subject to the above corrections, the normalized $q$-th order
moment in its Star integral form is
\beql{evf}
F_q(\epsilon)
\equiv { \xi_q(\epsilon) \ov \xi_q^{\rm norm}(\epsilon) }
\simeq  { \left\langle \sum_i a^{[q-1]} \right\rangle   \ov
     \left\langle \sum_i \langle b \rangle^{q-1} \right\rangle} \,.
\eeq
As in the case of the traditional moments \req{cif}, one would
measure $F_q(\epsilon)$ as a function of decreasing
$\epsilon$; a straight line in a plot of $\ln F_q$ vs.\
$\ln\epsilon$ would, as before, be interpreted as scaling
behavior of $\rho_q$. We caution, however, that there are important
issues which must be addressed before such conclusions can
be drawn, among them normalization effects.

The following points should be noted.
\begin{enumerate}
\item
Theoretical models are easily compared to data obtained with
the Star integral: when they predict the single-particle
distribution $\rho_1$ and correlation functions
$\rho_q$, their corresponding
Star integral moment is just given by the analytic integral
expression \req{cigc}.
Monte Carlo simulations should, of course, take into account
the proper removal of bias induced by finite sample size.
\item
The measurement of the Star integral is very economical.
Just as the numerator $\xi_q$
can be measured in an algorithm of order $N^2$ (the number of
particles per event), so the denominator requires only an algorithm
of the order of the square of the sample size, $\Nev^2$, a large
savings over the previous GHP and Snake integral algorithms
\ct{Lip92a}. (This savings is not destroyed by the
abovementioned corrections to obtain an unbiased normalization.)
When the order of the events in the sample is
random, this can be reduced even further by taking for the
``inner'' event average $\langle b \rangle$ only a fraction
of the full sample, e.g.\ $b \in \{a{-}1, a{-}2,\ldots, a{-}A\}$.
The smaller $A$, the faster the calculation but the larger the
statistical error. Since a small value of $A$ introduces a
considerable bias, which disappears as it is increased,
great care must be taken that $A$ is of a size where the normalization
becomes independent of its exact value. An optimal value for $A$
can be found for a given sample and length $\epsilon$ by
experimentation. Whenever doubt arises, the full $A = \Nev{-}1$
sum should be taken as this is exact.

For smaller values of $A$, the error in the denominator will be
nonnegligible and should be combined with the numerator error,
including covariances.
\item
For $\epsilon = \Delta x$, all theta functions
become unity by default and so
$F_q(\Delta x) = \langle N^{[q]} \rangle / \langle N \rangle^q$,
which is unity only when the event multiplicities in
$\Delta \bm{x}$ are Poisson-distributed. Furthermore,
when the distribution is purely random for a given $\epsilon$,
then $F_q(\epsilon)$ becomes unity;
see the discussion around Eq.\ \req{cusa} below.
\item
Annoying boundary effects due to the finiteness of
$\Delta\bm{x}$ are largely cancelled out, because the
vertical normalization used means that sphere counts
for centers $\bm{X}_i$ close to the boundary are reduced for
both numerator and denominator. In the business of galaxy
distributions, this remains a much-discussed topic \ct{Col92a};
since unfortunately only a single event exists in this
case, the horizontal normalization, which is vulnerable
to such boundary effects, has to be used there.
\item
Nevertheless, there may be instances where the horizontal normalization
may be preferred. The definition of fractal dimensions, for example,
is often couched in terms of the horizontally-normalized correlation
integral.
\item
The vertical normalization has the additional
advantage that the effects of the single-particle distribution
$\rho_1(\bm{x})$ are cancelled out to some
extent. While in many cases $\rho_1$ is constant (``stationary'')
or varying only weakly, it may in high energy collisions vary
quite strongly; especially the transverse momentum distribution
$\rho_1(p_\perp)$ is strongly peaked and then falls off exponentially.
Since sphere counts in both numerator and denominator vary about
equally as a function of $\rho_1$,
the trivial effects on $F_q$ of a strongly varying
single particle distribution are compensated.
However, this compensation is only partial: when $\epsilon$ is
small, each small domain is approximately flat and the cancellation
is fairly reliable. When $\epsilon$ is large, though,
the variation of a nonstationary $\rho_1$ within the
domain is averaged out before the ratio is taken instead
of the better reverse order. Ideally, one would divide up such
large domains into many small ones and sum up the contributions
only after normalization, but this is usually made impossible
by bad statistics. Also, even the so-called vertical factorial
moments \req{cif} used traditionally suffer for large bins
from the same effect, so that this problem can be ignored at the
present level of sophistication.
\item
Of greater concern is the possibility that the integration
of the center-coordinate $\bm{R}$ over the entire domain
$\Delta\bm{x}$ in Eqs.\ \req{cila} and \req{evb}
suffers from the same problem for any value of $\epsilon$.
For $\rho_1$'s that are weakly varying this does not matter very
much, but when they change drastically (such as $\rho_1(p_\perp)$)
a more vertical form should be used.
At some cost in CPU time, this can be implemented
straightforwardly. Instead of letting $\bm{R}$ range over the
entire space, we introduce discrete binning; for the Floating Box
\beql{noe}
\int_{\Delta\bm{x}} d\bm{R} \longrightarrow
\prod_{f=1}^{\sfd} \left(
{1\ov L} \sum_{\ell_f=1}^L
 \int_{(\ell_f-1)\delta R}^{\ell_f \delta R} dR_f   \right) ,
\eeq
with $\delta R = \Delta x/L$,
while the Floating Sphere of Eq.\ \req{cilf} yields,
after integrating out the angles,
\beql{nof}
\int_{\Delta\bm{x}} d\bm{R} \longrightarrow
{1\ov L} \sum_{\ell=1}^L \int_{(\ell-1)\delta R}^{\ell\delta R}
R^{{\sf d}-1}\, dR \;.
\eeq
By splitting the $\bm{R}$ integration in this way, the
correlation integral can be made largely independent of
the shape of $\rho_1$.
As one is usually interested in $F_q$ as a function of $\epsilon$
only, the number of subdivisions $L$ and bin sizes $\delta R$
can be kept fixed throughout a computation; the only requirement
is that the number of subdivisions should be large enough to
ensure that the final
correlation integral depends on $L$ only weakly.

For systems where $\rho_1$ varies strongly only with one
of its variables while varying weakly with the others,
it may be possible to define a ``hybrid'' correlation integral
with mixed normalization, implementing an $R$-sum as in
Eq.\ \req{noe} for this component only and not for the
weakly-varying components. This would be useful to boost
the number of counts.

As discussed in Section \ref{sec:sta}, splitting up $\bm{R}$
into different regions may in itself be useful in isolating
different physical effects. No sums over $\ell$ would be taken
for such cases.

\item
There remains the question how the traditional factorial moments
\req{cif} and Star correlation integrals \req{evf}
are to be compared. The simplest answer is that they should not
be compared at all, since by their different definitions they
cannot be expected to yield exactly the same results even for the
same data sample. Under ideal circumstances ($\Nev{\to}\infty$,
weakly varying $\rho_1$, etc.), the two should yield a similar
slope asymptotically. It may be helpful to
compare the two at equal $\epsilon$, but deviations should not be
interpreted as revealing anything fundamental. If one finds
large differences between traditional factorial moments \req{cif}
and correlation integrals, then they are mostly due to the
different normalization procedures used.

One can, however, check for consistency between
the BP factorial moments and the Star integrals. Both should
be the same for $\epsilon = \Delta x$. Also, for integer $M$,
one can do the Star integral {\it separately}
for every bin, with $\epsilon$ set to $\Delta x/M$, and then
add up the contributions. This sum over Star integrals should
then be identical to the vertical factorial moment $F_q^v(M)$.
\end{enumerate}

\newpage

\section{Cumulants}
\label{sec:cum}

\subsection{Definition and use}

The measurement of factorial moments and/or correlation integrals
may  be useful in itself in searching for a power law in the
correlation function. Moments do not, however, reveal the
true correlations, because the correlation function contains
uncorrelated parts which have to be subtracted.
This becomes clear on considering the effect that statistical
independence has on the correlation function.
Statistical independence of two points
$\bm{x}_1$ and $\bm{x}_2$ means that the correlation function
$\rho_2(\bm{x}_1,\bm{x}_2)$
factorizes into a product
$\rho_1(\bm{x}_1)\rho_1(\bm{x}_2)$.
Similarly, when $\bm{x}_1$ becomes statistically
independent of all other points $\bm{x}_{k\ne 1}$,
the higher order correlation functions factorize accordingly:
$\rho_q(\bm{x}_1,\bm{x}_2,\ldots,\bm{x}_q) \to
\rho_1(\bm{x}_1)\rho_{q-1}(\bm{x}_2,
\ldots,\bm{x}_q)$. All possible combinations of such
factorizations have to be subtracted from the original correlation
function before one can speak of the ``true'' correlations.

These reduced quantities, known as cumulants, are basic to
statistical analysis of any sort \ct{Stu87a}. They are constructed
precisely in such a way as to become zero whenever any one or more
of the points $x_k$ becomes statistically independent of the others.
(The often-used factorization $\rho_q \to \rho_1^q$
is only the most drastic form of statistical independence,
assuming that every point becomes independent of every other.)
Cumulants of different distributions are also additive under
convolution of the distributions \ct{Car87a} as well as being
invariant under change of origin \ct{Stu87a}.

The first few cumulants $C_q$ are, in terms of the correlation functions,
\beqa
\lleq{cua}
C_2(\bm{x}_1,\bm{x}_2) &=& \rho_2(\bm{x}_1,\bm{x}_2) -
\rho_1(\bm{x}_1)\rho_1(\bm{x}_2) \,, \\
\lleq{cub}
C_3(\bm{x}_1,\bm{x}_2,\bm{x}_3) &=& \rho_3(\bm{x}_1,\bm{x}_2,\bm{x}_3)
               -\ \rho_1(\bm{x}_1)\rho_2(\bm{x}_2,\bm{x}_3)\non\\
    & &\mbox{} -\ \rho_1(\bm{x}_2)\rho_2(\bm{x}_3,\bm{x}_1)
               -\ \rho_1(\bm{x}_3)\rho_2(\bm{x}_1,\bm{x}_2)\non\\
 & &\mbox{} +\ 2\,\rho_1(\bm{x}_1)\rho_1(\bm{x}_2)\rho_1(\bm{x}_3)\ ,\\
\lleq{cuc}
C_4(\bm{x}_1,\bm{x}_2,\bm{x}_3,\bm{x}_4)
 &=& \rho_4(\bm{x}_1,\bm{x}_2,\bm{x}_3,\bm{x}_4)
            - \sum_{(4)}
\rho_1(\bm{x}_1)\rho_3(\bm{x}_2,\bm{x}_3,\bm{x}_4)\non\\
 & &\mbox{} -\ \sum_{(3)}
\rho_2(\bm{x}_1,\bm{x}_2)\rho_2(\bm{x}_3,\bm{x}_4)\non\\
 & &\mbox{} +\ 2\,\sum_{(6)}
          \rho_1(\bm{x}_1)\rho_1(\bm{x}_2)\rho_2(\bm{x}_3,\bm{x}_4)\non\\
 & &\mbox{} -\ 6\,
\rho_1(\bm{x}_1)\rho_1(\bm{x}_2)\rho_1(\bm{x}_3)\rho_1(\bm{x}_4) \,.
\eeqa
The bracketed numbers under the sums
indicate the number of permutations of the arguments $\bm{x}_k$ which
have to be included. Further, omitting the arguments,
\beqa
\lleq{cuca}
C_5 &=& \rho_5 - \sum_{(5)}\rho_1\rho_4 - \sum_{(10)}\rho_2\rho_3
\non\\
& &\mbox{} + 2\sum_{(10)}\rho_1\rho_1\rho_3
    + 2\sum_{(15)}\rho_1\rho_2\rho_2
\non\\
& &\mbox{} - 6 \sum_{(10)}\rho_1\rho_1\rho_1\rho_2
    + 24 \rho_1^5 \,,
\eeqa
and so on for higher orders.

These equations have been utilized to find simple relations
between the vertical factorial moments $F_q$ of Eq.\req{cif}
and the integrated normalized cumulants \ct{Car90d}.
With ${\sf m} = (m_1,\ldots,m_{\sfd})$ as usual, the integrated
normalized cumulant is defined as
\beql{cud}
K_q^v(M) \equiv {1 \ov M^{\sfd}}  \sum_{\sf m}
{
\int_{\Omega_{\sf m}} \prod_k d\bm{x}_k\,
C_q(\bm{x}_1,\ldots,\bm{x}_q)
\ov
\left[ \int_{\Omega_{\sf m}} d\bm{x} \rho_1(\bm{x})
\right]^q
}     \,,
\eeq
which yields relations such as
$K_2^v = F_2^v - 1$, $K_3^v = F_3^v - 3 F_2^v + 2$ etc.
which can thus be utilized directly by experimentalists.
These relations hold exactly for $q \leq 3$ and approximately
for $q > 3$.
They are not true for horizontally normalized moments.

\subsection{Correlation integral cumulants}
\label{sec:cicu}

In contrast to Eq.\req{cud},
the correlation integral cumulant is defined as the integral
of $C_q$ over the domains $\Omega_S$ of Eq.\ \req{cilf} or
$\Omega_B$ of Eq.\ \req{ciqa} after appropriate transformation
to relative coordinates,
\beql{cuh}
K_q(\epsilon) \equiv
{ \rmf_q(\epsilon) \over \xi_q^{\rm norm}(\epsilon) } \,,
\eeq
with
\beqa
\lleq{cui}
\rmf_q(\epsilon) &\equiv&
\int_{\Delta \bm{x}} d\bm{R}
\int_0^\epsilon \prod_k d\bm{r}_k \,
C_q(\bm{R},\bm{r}_2,\ldots,\bm{r}_q) \non\\
&=& \int C_q(\bm{x}_1,\ldots,\bm{x}_q) \,
\Theta_{12}\Theta_{13}\ldots \Theta_{1q} \,
d\bm{x}_1\ldots d\bm{x}_q
\eeqa
the unnormalized factorial cumulant. The latter can be written
entirely in terms of the sphere counts introduced previously,
\beqa
a &=& \sum_{j}\Theta(\epsilon - X_{ij}^{aa})
   = \hat n(\bm{X}_i^a,\epsilon), \ \ \ \  j\neq i \non\\
\lleq{cuj}
b &=& \sum_j \Theta(\epsilon - X_{ij}^{ab})
   = \hat n_b(\bm{X}_i^a,\epsilon) \,;
\eeqa
(see Figures 2 and 3 for terms contributing to $a$ and $b$
respectively). To demonstrate this, we start with $q = 2$. Here
$\rmf_2 = \int (\rho_2 - \rho_1\rho_1)$, which by Eqs.\
\req{ciac}--\req{ciae} and \req{evb}--\req{eve} is seen to
yield (henceforth we suppress the dependence on $\epsilon$),
\beql{cuk}
\rmf_2 = \left\langle \sum_i ( a - \langle b \rangle ) \right\rangle .
\eeq
For $q=3$, the first term in the expansion \req{cub} of $C_3$
is just $\xi_3 = \langle \sum_i a^{[2]} \rangle$,
while the last term $\rho_1\rho_1\rho_1$ yields
$\xi_3^{\rm norm} = \langle\sum_i \langle b\rangle^2 \rangle$.
The three ``mixed terms'' involving both $\rho_2$ and
$\rho_1$ must be worked out explicitly. On the one hand,
\beqa
\lleq{cul}
\int d\bm{R} \int_0^\epsilon d\bm{r}_2\, d\bm{r}_3
\rho_1(\bm{x}_1) \rho_2(\bm{x}_2,\bm{x}_3)
&=& \Nev^{-2} \sum_{a,b} \sum_{i}\sum_{j\neq k}
  \Theta(\epsilon - X_{ij}^{ab})
  \Theta(\epsilon - X_{ik}^{ab}) \,, \non\\
&=& \left\langle \sum_i \langle b^{[2]} \rangle \right\rangle ,
\eeqa
while on the other hand, if $\bm{x}_1$ is contained in $\rho_2$,
\beql{cum}
\int d\bm{R} \int_0^\epsilon d\bm{r}_2\, d\bm{r}_3
\rho_1(\bm{x}_2) \rho_2(\bm{x}_1,\bm{x}_3)
=
\left\langle \sum_i a \langle b \rangle \right\rangle \,,
\eeq
so that, putting all the pieces together,
\beql{cun}
\rmf_3 = \left\langle \sum_i \left(
a^{[2]} - \langle b^{[2]} \rangle - 2 a \langle b \rangle
+ 2 \langle b \rangle^2 \right) \right\rangle .
\eeq
The constant recurrence of the outer event average and
$i$-sum suggests that we define an ``$i$-particle cumulant''
by
\beql{cuo}
\left\langle \sum_i \hat\rmf_q(i) \right\rangle \equiv \rmf_q \,,
\eeq
in terms of which we find
\beqa
\lleq{cupa}
\hat\rmf_2(i) &=& a - \langle b \rangle \,, \\
\lleq{cupb}
\hat\rmf_3(i)
&=& a^{[2]} - \langle b^{[2]} \rangle - 2 a \langle b \rangle
            + 2 \langle b \rangle^2 \,, \\
\lleq{cupc}
\hat\rmf_4(i)
&=&
a^{[3]} - \langle b^{[3]} \rangle
- 3 a^{[2]} \langle b \rangle - 3 a \langle b^{[2]} \rangle
  \non\\
& &\mbox{} + 6   \langle b \rangle \langle b^{[2]} \rangle
           + 6 a \langle b \rangle^2
           - 6 \langle b \rangle^3 \,, \\
\lleq{cupd}
\hat\rmf_5(i)
&=& a^{[4]} - \langle b^{[4]} \rangle
            - 4 a^{[3]} \langle b \rangle
            - 4 a \langle b^{[3]} \rangle
   \non\\
& &\mbox{}  -  6 a^{[2]} \langle b^{[2]} \rangle
            +  8 \langle b \rangle \langle b^{[3]} \rangle
            + 12 a^{[2]} \langle b \rangle^2
            +  6 \langle b^{[2]} \rangle \langle b^{[2]} \rangle
\non\\
& &\mbox{}  + 24 a \langle b \rangle \langle b^{[2]} \rangle
            - 36  \langle b \rangle^2 \langle b^{[2]} \rangle
            - 24 a \langle b \rangle^3
            + 24 \langle b \rangle^4 \,,
\eeqa
which are then summed over all particles $i$ and averaged
over all events to yield $\rmf_q$.

How these sums can be obtained graphically is illustrated  for
$q = 3$ in Figure 4(a) and $q=4$ in Figure 4(b).
The black squares represent individual
particles; those enclosed by a circle belong to the same event.
The center particle at $\bm{X}_i^a$ is connected
to the $q{-}1$ other particles by the lines representing the
theta functions. One now draws all possible event topologies with
$q{-}1$ lines connected to one center particle. For $p$ joining
lines within the $\bm{X}_i^a$ event, one
writes down a factor $a^{[p]}$; while lines connecting
$\bm{X}_i^a$ to $p$ particles in the same (other) event
yields a factor $\langle b^{[p]} \rangle$.  $p$ lines going to
different events $b$, $c,\ldots$ results in a factor
$\langle b \rangle^{p}$. Putting all such factors together
and assigning to each the appropriate sign and prefactor
from Eqs.\ \req{cua}{\it ff.}, one obtains the cumulant expansions
\req{cupa}--\req{cupd}.

Writing higher orders recursively in terms of
lower-order cumulants,
\beqa
\lleq{cuq}
\hat\rmf_3(i) &=& a^{[2]} - \langle b^{[2]} \rangle
               - 2 \langle b      \rangle \hat\rmf_2(i) \,,
\non\\
\hat\rmf_4(i) &=& a^{[3]} - \langle b^{[3]} \rangle
              - 3 \langle b       \rangle \hat\rmf_3(i)
              - 3 \langle b^{[2]} \rangle \hat\rmf_2(i) \,,
\non\\
\hat\rmf_5(i) &=& a^{[4]} - \langle b^{[4]} \rangle
              - 4 \langle b       \rangle \hat\rmf_4(i)
              - 6 \langle b^{[2]} \rangle \hat\rmf_3(i)
              - 4 \langle b^{[3]} \rangle \hat\rmf_2(i) \,,
\eeqa
we are led to the conjecture that
\beql{cur}
\hat\rmf_q(i) = a^{[q-1]} - \langle b^{[q-1]} \rangle
        - \sum_{p=2}^{q-1} {q-1 \choose p-1}
          \langle b^{[q-p]} \rangle \hat\rmf_p(i) \,,
\eeq
which, if proven for arbitrary $q\geq 5$, could open the way for
an easy calculation of cumulants to arbitrary order without doing
the messy algebra involved.

Just as the normalization $\xi_q^{\rm norm}$ must be corrected
for bias, we must for third and higher order cumulants also
correct for limited sample size. Here, too, using estimators to
correct for the effect of the limited-size sample
requires that all event sums go over unequal events. This again
results in corrections of the order $1/\Nev$, $1/\Nev^2$, $\ldots$
so that, for example, $\hat\rmf_3(i)$ above would acquire
the additional correction term
$2(\langle b \rangle^2 - \langle b^2 \rangle)/(\Nev - 2)$.
While these corrections can be quite important, we defer a discussion
of their origins and exact expressions to future work \ct{Liptbp}.

When an event sample has no correlations, the count around
a particle in event $a$ would on average be the same as when it
were inserted into other events,
\beql{cusz}
a^{[q-1]} \to \langle b^{[q-1]} \rangle
\eeq
which results in $\hat\rmf_q = 0$. This would of course be
true on average only. Less obvious
but also true is that if any one of the $q$ variables
of $C_q$ becomes
independent, the integrated cumulants $\rmf_q$ become zero also;
this can be shown graphically too.
A stronger condition of randomness, comparable to the Poisson
distribution in fixed bins, would be reflected by the behavior
\beqa
\lleq{cusa}
\langle \sum_i a^{[q-1]} \rangle &\longrightarrow&
\langle \sum_i \langle b \rangle^{q-1} \rangle , \non\\
\langle \sum_i \langle b^{[q-1]} \rangle \rangle &\longrightarrow&
\langle \sum_i \langle b \rangle^{q-1} \rangle ,
\eeqa
so that the moments would go to unity, $F_q \to 1$, for this case.

It must be stressed that
the $q$-th order cumulant contains no correlations of order
lower than $q$. Thus even if $\rmf_2 > 0$, $\rmf_3$ can still
be zero when there are no true third order correlations;
Eq.\ \req{cuq} is merely a convenient grouping of the terms.

We further see that, since only the basic quantities $a$ and $b$
are needed to construct cumulants, they can also be calculated
very economically with order $N^2$ algorithms.

The estimation of errors always provides a headache since mostly
one has to deal with the intricacies of error propagation and
covariances. In the Star integral formulation, however, this
process is much simplified; for the statistical error on $K_q$,
one merely has to calculate
\beql{cut}
\sigma^2(\rmf_q) =
{
\left\langle [ \sum_i \hat\rmf_q(i) ]^2 \right\rangle
- \rmf_q^2
\ov \Nev }
\eeq
and combine this with the corresponding statistical error for the
event mixing denominator. Errors for the
factorial moments $F_q$ and differentials of Section \ref{sec:diff}
are obtained with the same ease.

\subsection{Presenting cumulants}

While there may be many useful ways to plot correlations,
depending on what one is looking for, we recommend the following
format for cumulants. The second order cumulant
\beql{cuu}
K_2(\epsilon) =
{
\left\langle \sum_{i\neq j} \Theta(\epsilon - X_{ij}^{aa})
\right\rangle
\ov
\left\langle \sum_{i,j} \Theta(\epsilon - X_{ij}^{ab})
\right\rangle
}  - 1
\eeq
cannot be smaller than $-1$. Its theoretical maximum is
harder to find; but an estimate can be made by using the extreme
case where the events are extremely ``spiky'' but the spikes
are uniformly distributed in phase space (from event to event).
The event-averaged count will be approximately
$\bar n = \langle N \rangle (\epsilon/\Delta x)^{\sf d}$,
while the theta functions of the numerator are all unity
for our narrow spikes; this extreme case thus gives an approximate
upper limit
\beql{cuv}
-1 < K_2(\epsilon) <
{ \left\langle N(N-1) \right\rangle \ov
  \left\langle \bar n(\bar n - 1) \right\rangle  }
- 1 \simeq \left( \Delta x \ov \epsilon \right)^{2{\sf d}}  \,.
\eeq
to which a given sample $K_2$ can be compared. For higher orders,
one may then plot the $K_q$ directly and/or as the ratios
$K_q/K_2$, which would express $q$-th order correlations as a
fraction of second-order correlations. Testing for linking, on the
other hand \ct{Car90d}, one would plot the ratios
$K_q/K_2^{q-1}$.

Apart from calculating cumulants which are averaged over the
entire event sample, it may in specific cases be interesting to
look at ``single-event cumulants'' $\hat\rmf_q$ for rare events,
for example if a certain event or group of events exhibits unusual
patterns in a phase space plot. To see whether such cumulants differ
significantly from mere statistical noise in the
fluctuations, they should be plotted on top of the corresponding
event-averaged cumulant plus/minus twice (or three times) the error
$\rmf_q \pm 2\left(
\left\langle [ \sum_i \hat\rmf_q(i) ]^2 \right\rangle
- \rmf_q^2 \right)^{1/2}$.
Whether such single-event cumulants are a good idea will have to
be established in practice. We remind the reader that the
distribution of factorial moments is not necessarily gaussian
\ct{Lip91a} and that a large-deviation analysis
may also be appropriate in such cases  \ct{Sei93a}.

\section{Differential versions}
\label{sec:diff}

Correlation integrals and their cumulants described so far are
defined always in terms of a maximum distance $\epsilon$;
the ubiquitous theta functions ensure that all interparticle
distances $X_{ij}$ involved must be smaller than $\epsilon$.
The simplicity of this definition allows one to test clusters of many
particles at once, i.e. probe correlations of order 3 or higher,
something not possible using the conventional methods of
measuring correlation functions. It makes good sense, however,
to ask not only whether some interparticle distance is smaller
than some value, $X_{ij} < \epsilon$, but also whether it falls within
a certain {\it distance interval} $[\epsilon_{t-1},\epsilon_{t}]$.

To this purpose, we define a sequence of distances
$\epsilon_1, \epsilon_2, \ldots$ up to some maximum distance
determined by the total domain of integration.
This sequence can be either linear,
\beql{ddb}
\epsilon_t = t \epsilon_1 \,,
\eeq
or exponential,
\beql{ddbb}
\epsilon_t = \epsilon_1 c^{t-1} \,, \ \ c > 1,
\eeq
the second definition being useful because data presented as
function of $\ln\epsilon_t$ will be equally spaced.
These two sequences divide up the whole phase space
into adjacent and disjoint domains; for
${\sfd} = 1$ and $q = 2$, these domains are shown in
Figure 5 as the sequence of strips filling the entire domain.
We also introduce the {\it indicator function}
\beql{ddc}
I_t(X) \equiv \Theta(\epsilon_{t  } - X) -
\Theta(\epsilon_{t-1} - X) \,,
\eeq
which is unity when $\epsilon_{t-1} < X < \epsilon_t$ and zero
otherwise.

The differential forms are defined as follows (see also Figure 6).
Given a center
particle $\bm{X}_{i_1}^a$ in event $a$, the number of
particles situated a distance
$X_{i_1 i_k} \in [\epsilon_{t-1},\epsilon_t]$ away from
$\bm{X}_{i_1}$ is
\beql{dde}
\Delta\hat\xi_2(i,t) =
\hat n(\bm{X}_{i_1}^a,\epsilon_t) -
\hat n(\bm{X}_{i_1}^a,\epsilon_{t-1})
\equiv a_t - a_{t-1} \,,
\eeq
the latter defining the shortened notation we shall be using.
We next ask how many clusters of $q{-}1$ particles exist for
which the {\it maximum distance}
to $\bm{X}_{i_1}$ is in this interval,
$
\max(X_{i_1 i_2},\ldots,X_{i_1 i_q}) \in [\epsilon_{t-1},\epsilon_t]
$.
The answer is surprisingly simple: the number of such clusters is
\beql{ddf}
\Delta\hat\xi_q(i,t) =
\hat n(\bm{X}_{i_1}^a,\epsilon_t)^{[q-1]} -
\hat n(\bm{X}_{i_1}^a,\epsilon_{t-1})^{[q-1]}
\equiv a_t^{[q-1]} - a_{t-1}^{[q-1]} \,,
\eeq
since through use of
\beql{ddg}
\Theta(\epsilon_T - X) = \sum_{t=1}^T I_t(X)
\eeq
and
\beql{ddh}
\Theta(\epsilon - \max(X_{1},\ldots,X_{q} ))
= \prod_{p=1}^q \Theta(\epsilon - X_{p})
\eeq
we can show that
\beqa
\Delta\hat\xi_q(i_1,t) &=&
\left( \sum_{i_2} \Theta(\epsilon_t     - X_{i_1 i_2}) \right)^{[q-1]} -
\left( \sum_{i_2} \Theta(\epsilon_{t-1} - X_{i_1 i_2}) \right)^{[q-1]}
\non\\
\lleq{ddi}
&=& \sum_{i_2\neq \ldots \neq i_q}
I_t(\max(X_{i_1 i_2},\ldots,X_{i_1 i_q})) \,.
\eeqa
In Figure 6, one such cluster is shown, where at least one particle
besides the central one is in the shaded regions for the Floating
Sphere domain of Eq.\ \req{cilf}.

The normalization proceeds along the same lines as in Section
\ref{sec:evmix}, so that not surprisingly we find that the
normalized differential moment is
\beql{ddj}
\Delta F_q(t)
= {
\left\langle \sum_i  a_t^{[q-1]} - a_{t-1}^{[q-1]} \right\rangle
\ov
\left\langle \sum_i
\langle b_t     \rangle^{q-1}  -
\langle b_{t-1} \rangle^{q-1}  \right\rangle
} \,,
\eeq
while the cumulants are similarly
\beql{ddk}
\Delta K_q(t)
= {
\left\langle \sum_i  \hat\rmf_q(i,t) -
                      \hat\rmf_q(i,t{-}1) \right\rangle
\ov
\left\langle \sum_i
\langle b_t     \rangle^{q-1}  -
\langle b_{t-1} \rangle^{q-1}  \right\rangle
} \,.
\eeq
Both $\Delta F_q$ and $\Delta K_q$ are thus accessible to
measurement with
minimal additional effort. Usually, they will be plotted
as a function of $t$, i.e.\ the distance interval within which
the maximum interparticle distance would fall. $\Delta F_q$
can be thought of as the analogue to the ``factorial correlator''
defined in Ref.\ \ct{Bia86a} and $\Delta K_q$ to
their cumulants \ct{Egg91b}.

We conclude this section by listing the advantages of using the
differential moments and cumulants as a measure of cluster size.
\begin{itemize}
\item
First of all, they have all the advantages of the correlation
integrals in that their statistics will be higher and the data
points calculated more stable than the corresponding correlator.
Especially for higher dimensions, the large gain in statistics
will permit measurements that would be impossible otherwise.
\item
Differential moments and cumulants are almost immune to the
problems of converting from biased to unbiased estimators in the
normalization and in $\hat\rmf_q$. Because these corrections for
bias takes the form of an additive series, taking the difference
$\hat\rmf_q(i,t) - \hat\rmf_q(i,t{-}1)$ causes them
to cancel to a large degree.
\item
When a unique distance
can be defined as in Eq.\ \req{ciq}, the differential count is
unambiguous even in higher dimensions, unlike the
correlator, where distance definitions are ambiguous for
multidimensional analyses \ct{NA22-Ri}.
\item
Very importantly, the domains of integration of the normalized
differential moments and cumulants
are disjoint (Fig.\ 5), meaning that the data points will not be
correlated
amongst themselves, a constant bug in ordinary moments and
cumulants.
\item
A special status must be accorded to $\Delta F_2(t)$:
it behaves as a ``roaming distance'', looking for all particle
pairs that are a certain distance apart. This is suggestive of
interpreting $\Delta F_2$ as a kind of Fourier transform of
the distances \ct{Car91b}.
\end{itemize}

\section{Conclusion}
\label{sec:concl}

We have developed a general formalism for measuring correlations
of point distributions. The language used has been that of
high energy physics, but we believe that the method may be of
use in other fields also. The use of a delta function
notation has enabled us to derive the Star integral from
first principles and through its greater clarity
pointed to a number of important
extensions. Most important of these is that
the correlation integral used in the astronomy literature
\ct{Pal84a,Col92a} and suggested for high energy physics
\ct{Dre88a} appears to be in need of modification from
the form $\hat n^{q-1}$ to the ``factorial power'' form
$\hat n^{[q-1]}$. To assess whether such a modification is
possible or practical in galaxy distributions is beyond the scope
of this paper; for the limited number of particles encountered in
high energy physics, however, it seems an unavoidable and
clearly superior formulation.

The felicitous definition of relative coordinates leading to
the Star integral makes the latter very economical to
calculate, including all cumulants and differential quantities.
Since in addition the domains of integration are
the largest possible, we believe that the Star integral extracts
the maximum available correlation information from the data for
the minimum price in CPU time. In this
respect, it has proven superior to the traditional factorial
moments and older correlation measurements, especially when
correlations are measured in higher-dimensional spaces.
As computers continue their evolution to previously
unimaginable speeds and capacity, the present method should
become routine even for the higher orders.

Three issues have been dealt with only cursorily or not at all:
the question of eliminating the influence of the
total multiplicity distribution, the measurement of correlations
between different species of particles (e.g.\ distinguished
by their charge), and the problem of properly defining
fractal dimensions for nonstationary ensembles of events rather
than the usual single-event time series. We hope to return
to the latter in future.

The present paper has emphasized the {\it how to} rather than
the {\it what} of measurement. At first sight, it may seem
unnecessary to devote so much effort to the mere process of
measurement. However, high energy hadronic collisions, and to an
even greater extent nucleus-nucleus collisions, are presently in
a state where very few exact calculations on correlations
can be done and most theoretical work relies on assumptions which
often are hard to support or believe. In this context,
we believe that it is of cardinal importance that there should
be a clear and unambiguous method for analyzing correlations,
and that, if possible, a standard should be established by
which different experiments can be compared.

The confusion underlying the dynamics of correlations is
reflected in the cursory way in which we have treated the choice
of variables. While there are some theoretical preferences
\ct{And88a}, we are fairly ignorant of the
dynamics of the soft component and hence the best choice of
variables in this case. The original proposal regarding
self-similarity in particle production \ct{Bia86a} did not
have theoretical grounding in currently acceptable
physical theories but was based on a toy model to illustrate the
idea, while exact calculations of
correlations e.g.\ in a QCD framework \ct{Och92a} can
be applied only to high-$p_\perp$ processes.

The occurrence of different forms of correlation integrals might
cause unpleasant confusion among experimentalists who would prefer a
unique recipe to extract information of higher order
correlation functions. Unfortunately, there is a priori no
best choice. The different forms
merely reflect the freedom of choice of the particular {\it shape}
of the integration domain; but all commonly
probe the correlation functions by decreasing the {\it size\/} of the
integration domain. While the numerical values of the various
integrals may differ, the functional dependence is supposed
to be similar (this was shown numerically for the Snake and GHP
integrals \ct{Lip92a}).
Moreover, by suitable normalization most of the numerical
differences between the various forms can be divided out, so that
the choice of a particular form can be guided by practical
efficiency arguments.

On very general grounds, the choice of relative coordinates seems
a wise one, in whatever variables one may prefer. This is true
especially for cases where there is some degree of stationarity in
the distribution (i.e.\ invariance under translation),
which is generally assumed to be true for galaxy distributions and
perhaps for pion distributions at higher collision energies.
The ubiquitous use of such stationarity assumptions testifies to
their popularity. Not least, measurements in Bose-Einstein
interferometry rely on relative coordinates, whether in
their three-vector form $\bm{q} = \bm{p}_1 - \bm{p}_2$
or as a function of $Q^2 = -(p_1 - p_2)^2$.

In this context, we have aimed to provide a framework that is
adaptable to any future choice of variables and dynamical
theories. This will hopefully allow for clean measurements to
guide theoretical thinking, while remaining flexible in its
implementation.

\vspace{0.5cm}
\noindent {\bf Acknowledgements\/}:
HCE and PC gratefully acknowledge support by the Alexander von
Humboldt Foundation.
HCE and PL thank the Institut f\"ur Hochenergiephysik der \"OAW,
and the Theory Groups at the University of Arizona and
the University of Regensburg respectively for support.
PC acknowledges the hospitality of the GSI and the University of
Frankfurt.
Special thanks to the EMC and NA35 groups in Munich for insisting on
clear thinking, and to F.\ Botterweck, M.\ Charlet and
D.\ Weselka for useful discussions.
This work was supported in part by the US Department of Energy,
High Energy and Nuclear Physics Divisions, grant no.\
DE--FG02--88ER40456.


\vspace{20mm}

\newpage

\noindent
{\Large\bf Figure captions}

\vspace{6mm}
\noindent
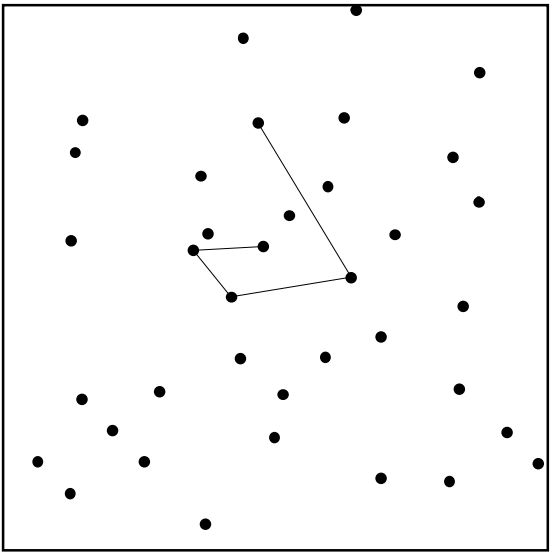
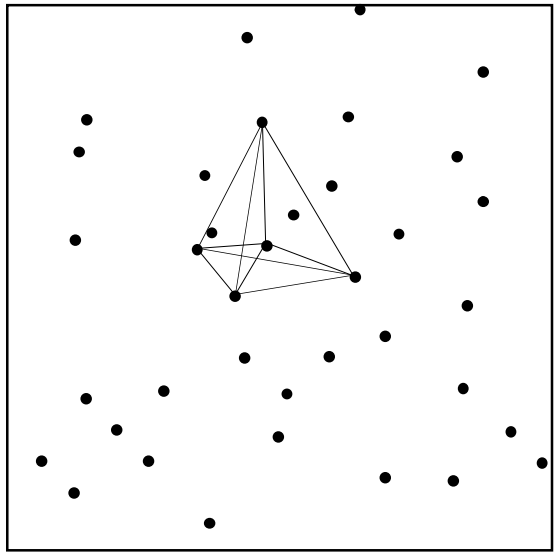
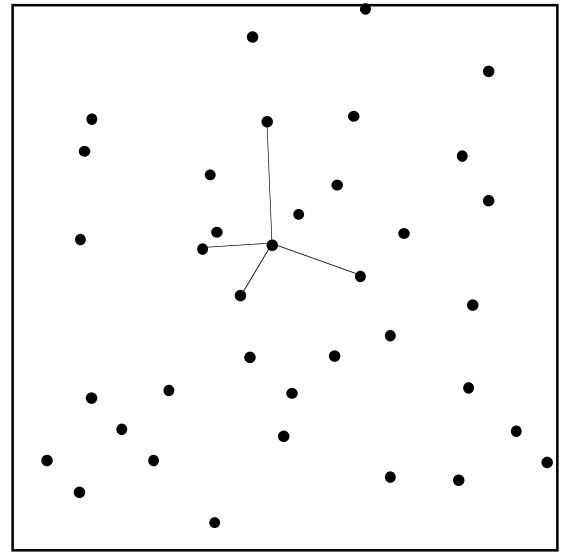
FIG. 1.
Different version of the correlation integral: (a) Snake,
(b) GHP, (c) Star. For a given set of $q$ particles in phase
space (here $q=5$) taken from the $N$ particles of a particular
event, pairwise distances are tested according to the
topology of joining lines shown. The longest of the lines
characterizes the size of the given $q$-tuple in every prescription.
The Star count is much more efficient
than the other prescriptions, see Section \ref{sec:cori}.

\vspace{6mm}
\noindent
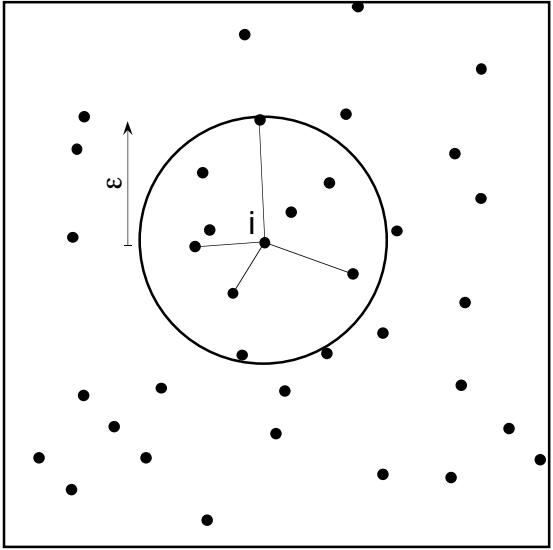
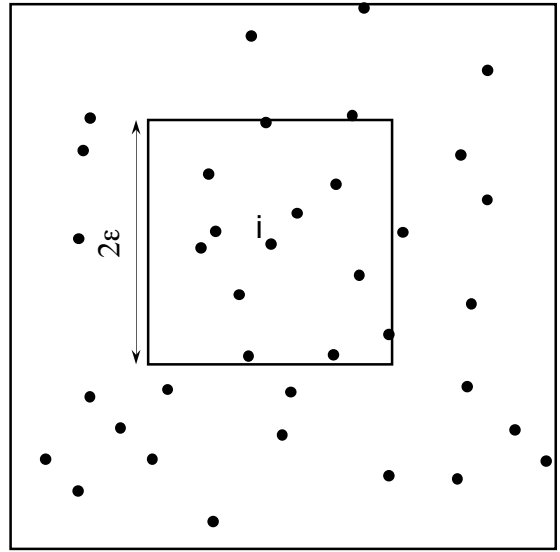
FIG. 2.
The conceptual advantage of the Star integral over other versions
stems from the fact that counting $q$-stars
can be reduced to computing factorials of {\it sphere counts}
$\hat n(\bm{X}_i,\epsilon)$.
(a)
``Floating Sphere'':
$\hat n(\bm{X}_i,\epsilon) =$ number of neighboring particles
within a sphere of radius $\epsilon$
centered at particle $i$ with coordinates $\bm{X}_i$.
The center particle itself is not included in the sphere count.
(b)
When the coordinates $\bm{x}$ have very different physical
properties in their components (such as $y, \phi$ and $\pt$),
the ``Floating Box'' may be a better choice as it treats
the distances along different coordinate axes independently.

\vspace{6mm}
\noindent
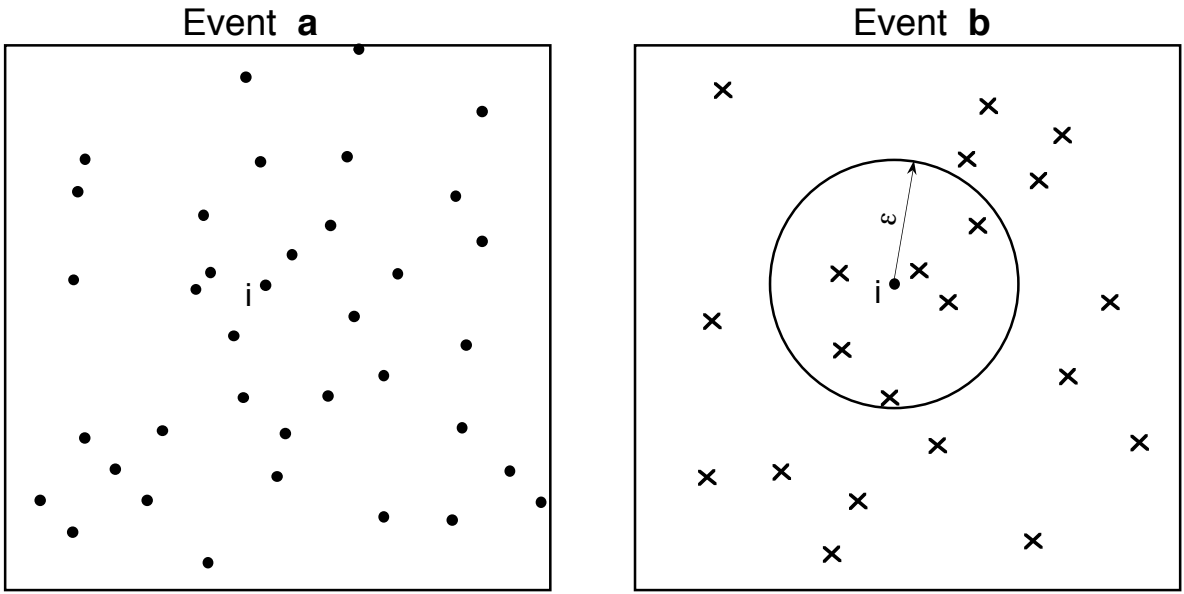
FIG. 3.
The basic building blocks for computing the normalization of
correlation integrals (Eqs.\ \req{evb}--\req{eve}) as well as
cumulants Eqs.\ \req{cuj}{\it ff.}) is the sphere count
$\hat n_b(\bm{X}^a_i,\epsilon)$.
While similar to the count of Fig.\ 2(a),
$\hat n_b(\bm{X}_i^a,\epsilon)$ performs a sphere count around
particle $\bm{X}_i^a$ taken from event $a$  (shown as a dot)
by placing it in the event $b$ and counting the $b$-particles
(shown as crosses) within the sphere.
In the example shown, $\hat n_b(\bm{X}_i^a,\epsilon) = 6$.
Averages over many events $b$ are taken while $\bm{X}_i^a$
is kept fixed.

\vspace{6mm}
\noindent
FIG. 4.
Schematic representation of event mixing terms entering
the (a) third  (b) fourth order cumulant
(Eqs.\ \req{cupb}--\req{cupc}).
For a given center particle $i$
in event $a$, the other particles in the $q$-tuple  can be
either within the same event $a$ or in different events
$b,c,\ldots$ (see text). $p$ particles in event $a$ lead to
a factor $a^{[p-1]}$; $p$ particles in event $b$ give a factor
$\langle b^{[p]} \rangle$ and a particle in $p$ different events
$b$, $c$, \ldots, gives a factor $\langle b \rangle^p$.
With the appropriate combinatorial prefactors, the sum of these
terms yields the third and fourth order cumulant integrals.

\vspace{6mm}
\noindent
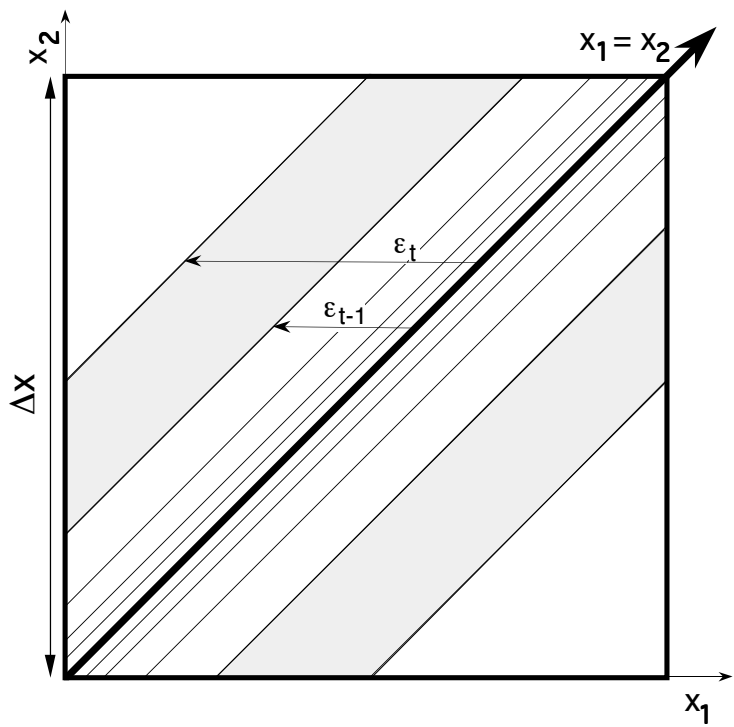
FIG. 5.
The exponential sequence of distances $\epsilon_t$ $(t=1,2,\ldots)$
of Eq.\ \req{ddbb}, used to define the differential
forms of correlation integrals of Section \ref{sec:diff}. The
shaded regions represent the integration area of the
differential integral $\Delta F_2(t)$ over the two-particle density
$\rho(x_1,x_2)$. Note that the coordinates $x$ in this figure are
one--dimensional (${\sf d} = 1$) and the labels 1 and 2 refer to
two different particles within the interval $\Delta x$.

\vspace{6mm}
\noindent
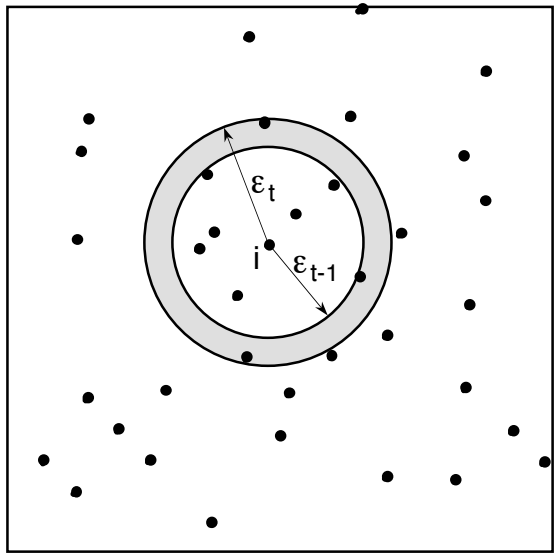
FIG. 6.
An example of the differential sphere count within the shaded area
in two dimensions (${\sf d} = 2$) gives
$\Delta \hat\xi_2(i,t)=a_t - a_{t-1} = 9-6 = 3$ (Eq.\ \req{dde}).
For higher orders,
the number of all $q$-stars with size within the interval
$[\epsilon_{t-1},\epsilon_t]$, i.e.\ with at least one
of the $q{-}1$ neighboring particles within the shaded area, is given
by $\Delta \hat\xi_q(i,t)=a_t^{[q-1]} - a_{t-1}^{[q-1]}$ of
Eq.\ \req{ddf}. In the present figure, for example,
$\Delta \hat\xi_3(i,t) = 42$.

\end{document}